\documentclass[usegraphicx]{mn2e}

\title[The gravitational instability of clumpy discs at high redshift]
      {Larson's scaling laws, and
       the gravitational instability of clumpy discs at high redshift}

\author[A. B. Romeo and O. Agertz]
       {Alessandro B. Romeo$^{1}$\thanks{E-mail: romeo@chalmers.se}
        and Oscar Agertz$^{2}$\\
        $^{1}$Department of Earth and Space Sciences,
              Chalmers University of Technology,
              SE-41296 Gothenburg, Sweden\\
        $^{2}$Department of Physics,
              University of Surrey,
              Guildford GU2 7XH, United Kingdom}

\begin{document}

\date{Accepted 2014 May 12.
      Received 2014 May 2; in original form 2014 March 4}

\pagerange{\pageref{firstpage}--\pageref{lastpage}}

\pubyear{2014}

\maketitle

\label{firstpage}

\begin{abstract}
Gravitational instabilities play a primary role in shaping the clumpy
structure and powering the star formation activity of gas-rich high-redshift
galaxies.  Here we analyse the stability of such systems, focusing on the
size and mass ranges of unstable regions in the disc.  Our analysis takes
into account the mass-size and linewidth-size scaling relations observed in
molecular gas, originally discovered by Larson.  We show that such relations
can have a strong impact on the size and mass of star-forming clumps, as well
as on the stability properties of the disc at all observable scales, making
the classical Toomre parameter a highly unreliable indicator of gravitational
instability.  For instance, a disc with $Q=1$ can be far from marginal
instability, while a disc with $Q\ll1$ can be marginally unstable.  Our work
raises an important caveat: if clumpy discs at high redshift have
scale-dependent surface densities and velocity dispersions, as implied by the
observed clump scaling relations, then we cannot thoroughly understand their
stability and star formation properties unless we perform multi-scale
observations.  This will soon be possible thanks to dedicated ALMA surveys,
which will explore the physical properties of super-giant molecular clouds at
the peak of cosmic star formation and beyond.
\end{abstract}


\section{INTRODUCTION}

Today it is well established that the majority of the stellar mass
observed in galaxies formed at high redshift, and in particular that
the mean cosmological star-formation-rate density peaks at redshift
1--3 (e.g., Hopkins \& Beacom 2006).  Recent semi-empirical studies
(e.g., Behroozi et al.\ 2013) have allowed understanding how the peak
redshift of individual galaxies depends, on average, on the mass of
the host dark matter halo, with today's $L_{*}$ galaxy population
forming stars at peak efficiency around $z=1\mbox{--}2$.
Understanding the complex behaviour of galaxy assembly is a daunting
task for galaxy formation theory (e.g., Weinmann et al.\ 2011), and
highlights the need to build robust models that are capable of
predicting how star formation proceeds in high-redshift galaxies.

The morphology and star formation properties of massive high-redshift
galaxies are very different from those of present-day quiescent
spirals and ellipticals.  Extended clumpy irregular discs with
kpc-sized star-forming clumps as massive as
$M\sim10^{7}\mbox{--}10^{9}\,\mbox{M}_{\odot}$ are observed in the
Hubble Ultra Deep Field (UDF; e.g., Elmegreen et al.\ 2007, 2009), a
population that is rare today.  Multi-wavelength observational
evidence (e.g., Elmegreen \& Elmegreen 2006; Shapiro et al.\ 2008;
Tacconi et al.\ 2010) suggests that clumps generally form in gas-rich
spiral discs rather than in mergers, although the latter scenario
cannot be completely ruled out (e.g., Overzier et al.\ 2008).
Numerical work by Bournaud et al.\ (2007) and Elmegreen et al.\ (2008)
demonstrated that internal disc fragmentation can reproduce many of
the observables of clumpy high-redshift galaxies.  Using
high-resolution hydrodynamical simulations in a fully cosmological
framework, Agertz et al.\ (2009b) demonstrated that super-massive
clumps are a natural outcome of fragmenting massive gas-rich discs,
formed from multi-phase cosmological accretion (see also Ceverino et
al.\ 2010).

The size and mass of such clumps can be predicted using simple arguments, if
one assumes that the disc is marginally unstable according to Toomre's
stability criterion (e.g., Noguchi 1998, 1999; Dekel et al.\ 2009; Genzel et
al.\ 2011).  This assumption makes sense because current dynamical models of
high-redshift star-forming galaxies suggest that their discs are driven by
self-regulation processes, which keep them close to marginal instability
(e.g., Noguchi 1998, 1999; Agertz et al.\ 2009a; Dekel et al.\ 2009; Burkert
et al.\ 2010; Krumholz \& Burkert 2010; Cacciato et al.\ 2012; Forbes et
al.\ 2012, 2014).  If $Q\equiv\kappa\sigma/\pi G\Sigma=1$, then there is a
single unstable wavelength, $\lambda=2\sigma^{2}/G\Sigma$, and the associated
mass is $M\sim\Sigma\lambda^{2}=4\sigma^{4}/G^{2}\Sigma$.  In the gas disc of
the Milky Way, these quantities are comparable to the maximum size and mass
of giant molecular clouds, i.e.\ $\lambda\sim100\,\mbox{pc}$ and
$M\sim10^{6}\,\mbox{M}_{\odot}$ (see, e.g., Glazebrook 2013).  In
high-redshift discs, both the surface density $\Sigma$ and the velocity
dispersion $\sigma$ of molecular gas are typically one order of magnitude
larger than in the Milky Way (see again Glazebrook 2013).  As
$\lambda\propto\sigma^{2}/\Sigma$ and $M\propto\sigma^{4}/\Sigma$, we get
$\lambda\sim1\,\mbox{kpc}$ and $M\sim10^{9}\,\mbox{M}_{\odot}$, which are the
typical clump size and mass.  Genzel et al.\ (2011) and Wisnioski et
al.\ (2012) showed that the clumps are located in regions of the disc where
$Q\la1$.  This provides further evidence that in clumpy discs at high
redshift there is a strong link between star formation and gravitational
instability.

In spite of its predictive power, such a scenario neglects an important
aspect of the problem: in clumpy discs, the surface density and velocity
dispersion depend on the size of the region over which they are measured
(Romeo et al.\ 2010; Hoffmann \& Romeo 2012), contrary to what is generally
assumed (see, e.g., Glazebrook 2013).  In fact, there is mounting evidence
that molecular gas is characterized by mass-size and linewidth-size scaling
relations:
\begin{equation}
\Sigma\propto\ell^{a},
\;\;\;\;\;\mathrm{i.e.}\ M\propto\ell^{2+a},
\end{equation}
\begin{equation}
\sigma\propto\ell^{b},
\end{equation}
where $\Sigma$ and $M$ are the mass column density and the mass of the clump,
$\sigma$ is its 1D velocity dispersion, and $\ell$ is the clump size.
\begin{enumerate}
\item The most compelling evidence of such a link comes from observations of
  molecular clouds in the Milky Way and nearby galaxies (see, e.g.,
  Hennebelle \& Falgarone 2012, and references therein; Donovan Meyer et
  al.\ 2013; Kauffmann et al.\ 2013; Kritsuk et al.\ 2013; Kruijssen \&
  Longmore 2013).  These observations show that both Galactic and
  extragalactic molecular clouds are fairly well described by the so-called
  `\emph{Larson's scaling laws}', $a=0$ and $b=\frac{1}{2}$ (Larson 1981;
  Solomon et al.\ 1987), although the uncertainties are still significant:
  $-0.8\la a\la0.7$ (Beaumont et al.\ 2012), and $0.2\la b\la1.1$ (Shetty et
  al.\ 2012).
\item Similar scaling exponents are found in high-resolution simulations of
  molecular clouds and supersonic turbulence (see, e.g., Hennebelle \&
  Falgarone 2012, and references therein; Beaumont et al.\ 2013; Federrath
  2013; Kritsuk et al.\ 2013; Bertram et al.\ 2014; Fujimoto et al.\ 2014;
  Ward et al.\ 2014).  The latter simulations show that $a$ depends not only
  on the Mach number of the gas, but also on turbulence forcing (Federrath et
  al.\ 2009, 2010; Federrath 2013) and self-gravity (Collins et al.\ 2012;
  Kritsuk et al.\ 2013).  In contrast, at high Mach numbers, $b$ is
  approximately constant and close to 0.5 (see again Federrath 2013; Kritsuk
  et al.\ 2013).
\item Larson-type scaling relations have recently been observed, for the
  first time, in the dense star-forming clumps of a high-redshift galaxy: the
  strongly lensed sub-millimetre galaxy SMM J2135--0102 at $z=2.32$, also
  known as the cosmic eyelash (Swinbank et al.\ 2011).  Although this is the
  only detection of super-giant molecular clouds at high redshift, it will
  soon be followed by many such observations, which will exploit the
  unprecedented resolution and sensitivity of the Atacama Large
  Millimeter/submillimeter Array (ALMA) for exploring the physical properties
  of molecular gas at $z\ga2$ (see, e.g., Glazebrook 2013).
\end{enumerate}

Romeo et al.\ (2010) explored the gravitational instability of clumpy gas
discs, and showed that the mass-size and linewidth-size scaling relations of
the clumps can have a strong impact on disc instability.  For instance, they
can excite three main instability regimes, two of which have no classical
counterpart.  Hoffmann \& Romeo (2012) generalized this result to
two-component discs of clumpy gas and old stars, and analysed the stability
of spirals from The H\,\textsc{i} Nearby Galaxy Survey (THINGS).

In this paper, we investigate the gravitational instability of clumpy discs
at high redshift, focusing on the size and mass ranges of unstable regions
(see Sect.\ 2).  We begin by spelling out the assumptions of our stability
analysis and summarizing the results of Romeo et al.\ (2010), which are
fundamental to a proper understanding of this paper (see Sect.\ 2.1).  Next,
we discuss the effects of varying the clump scaling relations across the
observed ranges of $a$ and $b$, and illustrate how the spatial resolution
affects the \emph{inferred} stability properties of the disc, if the observed
$\Sigma$ and $\sigma$ are scale-dependent (see Sect.\ 2.2).  This is a
complex aspect of the problem, which should be taken into account when
analysing the stability of high-redshift star-forming galaxies.  Last but not
least, we discuss the properties of discs close to marginal instability (see
Sect.\ 2.3).  As pointed out above, this is the condition generally assumed
for estimating the typical size and mass of the clumps.  The disc scale
height is expected to play a significant role in this scenario, since it is
the scale at which galactic turbulence undergoes a transition from 3D to 2D
(e.g., Bournaud et al.\ 2010), and this may be accompanied by a break in the
clump scaling relations.  We discuss this aspect of the problem in Sect.\ 3.
The conclusions of our paper are drawn in Sect.\ 4.

\section{GRAVITATIONAL INSTABILITIES IN CLUMPY DISCS}

\subsection{The main instability regimes}

When analysing the stability of high-redshift star-forming galaxies, it is
generally assumed that the surface density of the disc is dominated by
molecular gas (g) and young stars ($\star$),
\begin{equation}
\Sigma=\Sigma_{\mathrm{g}}+\Sigma_{\star}\,,
\end{equation}
and that the gaseous and stellar components have similar kinematic
properties, so that the velocity dispersion of the disc is simply
\begin{equation}
\sigma=\sigma_{\mathrm{g}}=\sigma_{\star}
\end{equation}
(e.g., Burkert et al.\ 2010; Krumholz \& Burkert 2010; Puech 2010; Genzel et
al.\ 2014).  This assumption makes sense because the mass fraction of
molecular gas increases steeply with redshift (e.g., Daddi et al.\ 2010;
Tacconi et al.\ 2010; Carilli \& Walter 2013), and because most of the stars
in high-redshift discs were probably formed during the on-going starburst and
did not have time to heat up significantly (see again Burkert et al.\ 2010;
Krumholz \& Burkert 2010; Puech 2010; Genzel et al.\ 2014).  Older
generations of stars with
$\sigma_{\mathrm{old}\,\star}\gg\sigma_{\mathrm{g}}$ could also exist
(Glazebrook 2013), but they would play a negligible role in the gravitational
instability of the disc, even if
$\Sigma_{\mathrm{old}\,\star}\ga\Sigma_{\mathrm{g}}$, because the resulting
$Q$ stability parameter would still be dominated by the young gaseous-stellar
component (Romeo \& Falstad 2013).

As pointed out in Sect.\ 1, this simple model does not capture an important
aspect of the problem: in clumpy discs, the mass-size and linewidth-size
scaling relations of the clumps can have a strong impact on disc instability
(Romeo et al.\ 2010).  Here we take such relations into account, and assume
that molecular gas and young stars have similar scaling properties, so that
the surface density and the velocity dispersion of the disc are
scale-dependent and given by:
\begin{equation}
\Sigma(\ell)=\Sigma_{0}\left(\frac{\ell}{\ell_{0}}\right)^{a},
\;\;\;\;\;a=a_{\mathrm{g}}=a_{\star}\,;
\end{equation}
\begin{equation}
\sigma(\ell)=\sigma_{0}\left(\frac{\ell}{\ell_{0}}\right)^{b},
\;\;\;\;\;b=b_{\mathrm{g}}=b_{\star}\,.
\end{equation}
This assumption makes sense because newborn stars inherit the scaling
properties of the parent gas (e.g., Larson 1979; S\'{a}nchez et al.\ 2010).
Note that, since Eqs (5) and (6) are power laws, the choice of $\ell_{0}$ is
arbitrary.  What really matters is not $\ell_{0}$ itself but the values of
$A\equiv\Sigma_{0}/\ell_{0}^{a}$ and $B\equiv\sigma_{0}/\ell_{0}^{b}$, which
unfortunately are poorly constrained.  A physically meaningful choice would
be to identify $\ell_{0}$ with the disc scale height, $h$, which is the
natural smoothing scale of galactic discs (Romeo 1994).  However, $\Sigma$
and $\sigma$ are often measured at scales comparable to the disc scale
length, $R_{\mathrm{d}}$ (e.g., Puech 2010), or at intermediate scales (e.g.,
Genzel et al.\ 2014).  To make our analysis readily applicable, we choose to
identify $\ell_{0}$ with the scale at which $\Sigma$ and $\sigma$ are
measured, and assume that this is also the scale at which the Toomre
parameter and the 2D Jeans length are inferred:
\begin{equation}
Q_{0}=\frac{\kappa\sigma_{0}}{\pi G\Sigma_{0}}\,,
\end{equation}
\begin{equation}
L_{\mathrm{J}0}
\equiv\frac{2\pi}{k_{\mathrm{J}0}}
=\frac{\sigma_{0}^{2}}{G\Sigma_{0}}\,,
\end{equation}
where $k_{\mathrm{J}0}$ is the 2D Jeans wavenumber (once again, choosing
$\ell_{0}=h$ or $\ell_{0}=R_{\mathrm{d}}$ is only conceptually different from
our choice; the results are identical).  Hereafter we will refer to
$\ell_{0}$ as `the spatial resolution (scale)', like Leroy et al.\ (2008) and
Genzel et al.\ (2014).  This scale should \emph{not} be confused with the
resolution limit of the observations: $\Sigma$ and $\sigma$ are usually
measured averaging over scales larger than the beam size.  Note also that
$\ell_{0}$ cannot have any influence on the actual stability properties of
the disc.  However, since $\ell_{0}$ affects the \emph{inferred} values of
$Q$ and $L_{\mathrm{J}}$, $Q(\ell)=Q_{0}\,(\ell/\ell_{0})^{b-a}$ and
$L_{\mathrm{J}}(\ell)=L_{\mathrm{J}0}\,(\ell/\ell_{0})^{2b-a}$, it will also
have a significant effect on the \emph{derived} conditions for gravitational
instability.

As discussed above, clumpy discs at high redshift are dynamically similar to
gas discs with scale-dependent surface density and velocity dispersion.  The
gravitational instability of such discs was explored by Romeo et al.\ (2010).
Below we summarize some of their results, which are fundamental to a proper
understanding of Sects 2.2 and 2.3.

\begin{figure}
\includegraphics[scale=.98]{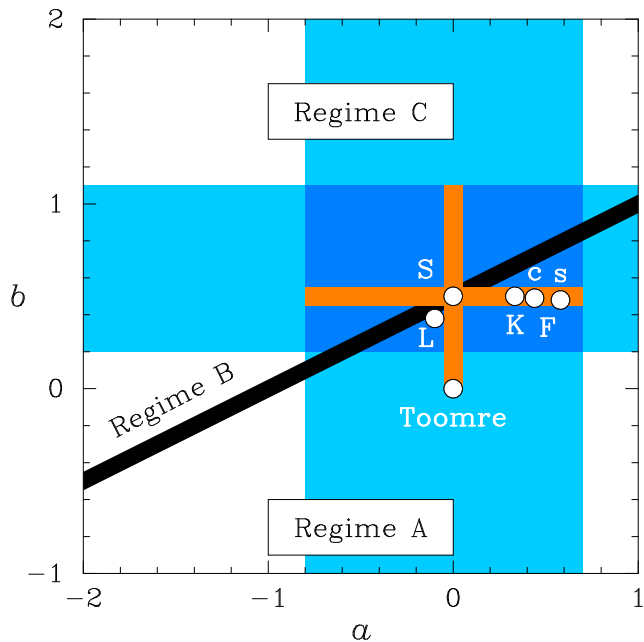}
\caption{The main instability regimes of clumpy discs.  The clumps are
  characterized by Larson-type scaling relations: $\Sigma\propto\ell^{a}$ and
  $\sigma\propto\ell^{b}$, where $\Sigma$ is the mass column density,
  $\sigma$ is the 1D velocity dispersion, and $\ell$ is the clump size.
  Regime B is a transition between Toomre-like instability (Regime A) and
  small-scale instability (Regime C).  Also shown are the ranges of $a$ and
  $b$ observed in molecular clouds (shaded), the ranges analysed in Sects 2.2
  and 2.3 (highlighted), and the specific values of $(a,b)$ further analysed
  in Sect.\ 2.3.}
\end{figure}

If the disc is subject to local axisymmetric perturbations, as is generally
assumed, then its response is described by a dispersion relation that we now
write as:
\begin{equation}
\frac{\omega^{2}}{\kappa^{2}}=1-
\frac{4}{Q_{0}^{2}}\,
\frac{\Sigma(\ell)/\Sigma_{0}}{(\ell/L_{\mathrm{J}0})}+
\frac{4}{Q_{0}^{2}}\,
\frac{\sigma^{2}(\ell)/\sigma_{0}^{2}}{(\ell/L_{\mathrm{J}0})^{2}}\,,
\end{equation}
where $\omega$ and $k=2\pi/\ell$ are the frequency and the wavenumber of the
perturbation, and $\kappa$ is the epicyclic frequency.  Note that Eq.\ (9)
applies to realistically thick discs (see sect.\ 2.1 of Romeo et al.\ 2010).
If the disc has volume density $\rho$ and scale height $h$, then
$\Sigma\approx2\rho\ell$ for $\ell\la h$ and $\Sigma\approx2\rho h$ for
$\ell\ga h$.  In both cases, the associated mass is $M\sim\Sigma\ell^{2}$.
The range $\ell\la h$ corresponds to the case of 3D turbulence (i.e.\ to the
usual clump scaling relations), whereas the range $\ell\ga h$ corresponds to
the case of 2D turbulence (i.e.\ to a large-scale extrapolation of the clump
scaling relations; a more realistic case will be discussed in Sect.\ 3).
Note also that Eq.\ (9) is a relation between $\omega^{2}/\kappa^{2}$ and
$\ell/L_{\mathrm{J}0}$, which is affected by four parameters: $Q_{0}$ (the
classical stability parameter), $\ell_{0}/L_{\mathrm{J}0}$ (a parameter that
couples gravitational instability with spatial resolution), $a$ and $b$ (the
logarithmic slopes of the clump scaling relations).  It turns out that $a$
and $b$ have an important effect on the shape of the dispersion relation, and
hence on the condition for gravitational instability ($\omega^{2}<0$).
Variations in the scaling properties of the clumps can drive high-redshift
discs across three main instability regimes.  Such regimes are illustrated in
Fig.\ 1, together with the ranges of $a$ and $b$ observed in molecular clouds
(shaded), and other useful information (which will be discussed in Sects 2.2
and 2.3).

\begin{figure*}
\includegraphics[scale=.99]{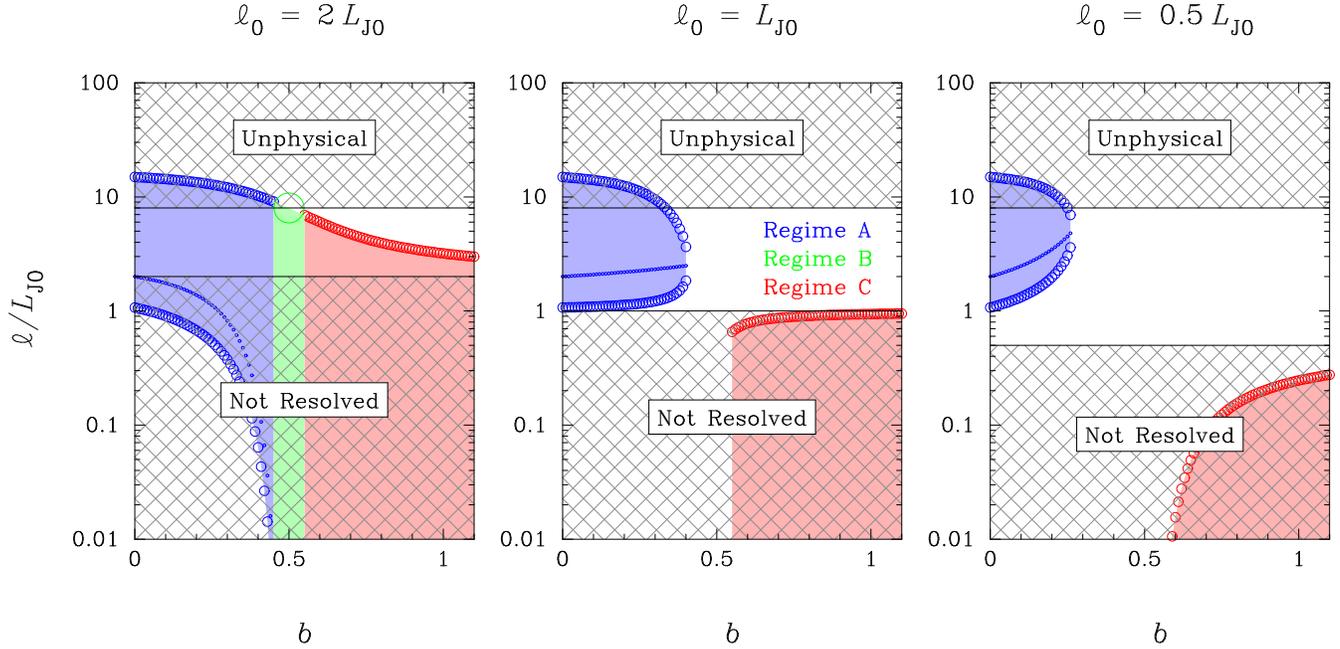}
\caption{Size range of unstable regions in clumpy discs: effect of varying
  the linewidth-size scaling relation of the clumps ($\sigma\propto\ell^{b}$)
  at different spatial resolutions ($\ell_{0}$).  We assume that the
  mass-size scaling relation is $M\propto\ell^{2}$ (Larson's third law), and
  that the Toomre parameter (at scale $\ell=\ell_{0}$) is
  $Q_{0}\equiv\kappa\sigma_{0}/\pi G\Sigma_{0}=0.5$.  Lengths are measured in
  units of the 2D Jeans length,
  $L_{\mathrm{J}0}\equiv\sigma_{0}^{2}/G\Sigma_{0}$.  Also shown is the most
  unstable scale for discs in Regime A (this quantity vanishes in Regimes B
  and C).}
\end{figure*}

\begin{figure*}
\includegraphics[scale=.99]{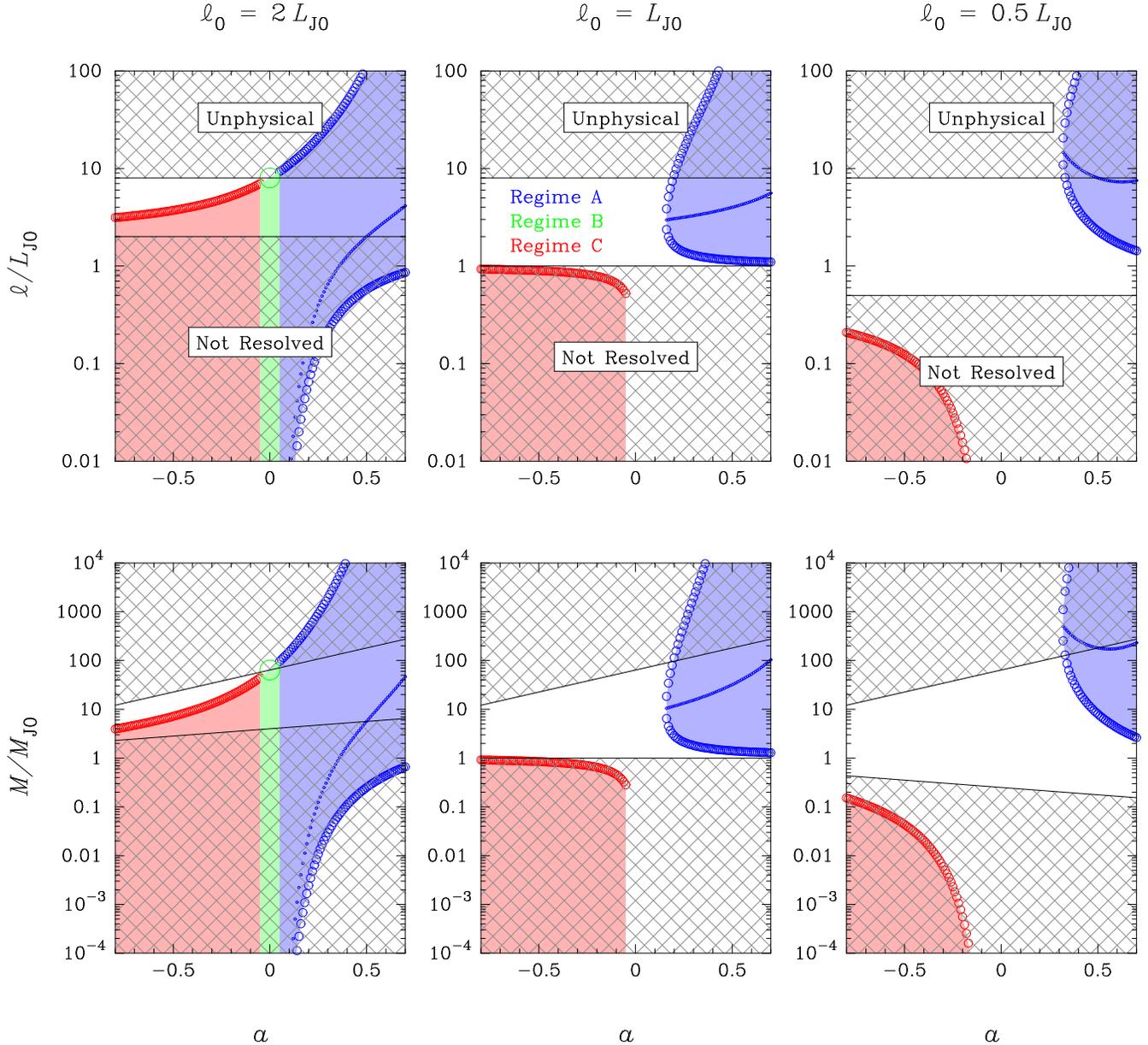}
\caption{Size range (top) and mass range (bottom) of unstable regions in
  clumpy discs: effect of varying the mass-size scaling relation of the
  clumps ($M\propto\ell^{2+a}$) at different spatial resolutions
  ($\ell_{0}$).  We assume that the linewidth-size scaling relation is
  $\sigma\propto\ell^{1/2}$ (Larson's first law), and that the Toomre
  parameter (at scale $\ell=\ell_{0}$) is $Q_{0}\equiv\kappa\sigma_{0}/\pi
  G\Sigma_{0}=0.5$.  Lengths and masses are measured in units of the 2D Jeans
  length, $L_{\mathrm{J}0}\equiv\sigma_{0}^{2}/G\Sigma_{0}$, and the
  associated mass,
  $M_{\mathrm{J}0}\equiv\Sigma(L_{\mathrm{J}0})\,L_{\mathrm{J}0}^{2}$.  Also
  shown are the most unstable scale and the associated mass for discs in
  Regime A (these quantities vanish in Regimes B and C).}
\end{figure*}

\begin{itemize}
\item In \emph{Regime A}, i.e.\ for $b<\frac{1}{2}\,(1+a)$ and $-2<a<1$, the
  stability of the disc is controlled by $Q_{0}$: the disc is stable at all
  scales if and only if $Q_{0}\geq\overline{Q}_{0}$, where the stability
  threshold $\overline{Q}_{0}$ depends on $a$, $b$ and
  $\ell_{0}/L_{\mathrm{J}0}$.
\item In \emph{Regime C}, i.e.\ for $b>\frac{1}{2}\,(1+a)$ and $-2<a<1$, the
  stability of the disc is no longer controlled by $Q_{0}$: the disc is
  always unstable at small scales (i.e.\ as
  $\ell/L_{\mathrm{J}0}\rightarrow0$) and stable at large scales (i.e.\ as
  $\ell/L_{\mathrm{J}0}\rightarrow\infty$).
\item In \emph{Regime B}, i.e.\ for $b=\frac{1}{2}\,(1+a)$ and $-2<a<1$, the
  disc is stable at all scales if and only if
  $\ell_{0}/L_{\mathrm{J}0}\leq1$.  This is a regime of transition between
  stability \`{a} la Toomre (Regime A) and instability at small scales
  (Regime C).  Thus even small variations in the scaling properties of the
  clumps can drive the disc into Regime A or Regime C, and have a strong
  impact on its gravitational instability.
\end{itemize}

Note that the stability criteria for Regimes A and B are
\emph{scale-invariant}, i.e.\ they only apparently depend on the spatial
resolution scale $\ell_{0}$: if the disc is stable at a given $\ell_{0}$,
then it will also be stable at any other spatial resolution.  Note also that
similar instability regimes are found in the Milky Way and nearby galaxies,
but at scales smaller than about 100 pc (Hoffmann \& Romeo 2012).

\subsection{Size and mass ranges of unstable regions}

Eq.\ (9) can be used not only for identifying the main instability regimes of
clumpy discs, but also for predicting the size and mass ranges of unstable
regions in such systems.  These correspond to the range(s) of $\ell$ where
$\omega^{2}<0$ and to the associated range(s) of
$M\equiv\Sigma(\ell)\,\ell^{2}\propto\ell^{2+a}$ (numerical factors are
irrelevant, since we are interested in mass ratios).  Here we focus on this
aspect of the problem, and analyse the effects of varying the clump scaling
relations across the ranges highlighted in Fig.\ 1.  The idea behind our
choices of $a$ and $b$ is to vary one parameter at a time, starting from
$a=0$ and $b=\frac{1}{2}$ (Larson's scaling laws), and spanning the ranges of
$a$ and $b$ observed in molecular clouds (shaded).  We vary $b$ down to $b=0$
so as to include the classical case of Toomre instability.  Concerning the
other parameters, we choose $\ell_{0}/L_{\mathrm{J}0}=0.5,1,2$ so as to
sample the (in)stability condition for discs governed by Larson's scaling
laws (Regime B), and $Q_{0}=0.5$ so as to represent the state of violent
gravitational instability observed in high-redshift star-forming galaxies
(e.g., Puech 2010; Genzel et al.\ 2014).  Note that these values of
$\ell_{0}/L_{\mathrm{J}0}$ and $Q_{0}$ match those found in the rings and
outer discs of SINS/zC-SINF galaxies at $z\sim2$ (Genzel et al.\ 2014).%
\footnote{The rings and outer discs of such galaxies typically have
  $\kappa\approx\mbox{20--80}\;\mbox{km\,s}^{-1}\,\mbox{kpc}^{-1}$,
  $\sigma\approx60\;\mbox{km\,s}^{-1}$,
  $\Sigma_{\mathrm{g}}\approx400\;\mbox{M}_{\odot}\,\mbox{pc}^{-2}$ and
  $\Sigma_{\star}\approx200\;\mbox{M}_{\odot}\,\mbox{pc}^{-2}$ (Genzel,
  private communication).  This yields $Q_{0}\approx\mbox{0.2--0.6}$ and
  $L_{\mathrm{J}0}\approx1.4\;\mbox{kpc}$.  The spatial resolution scale is
  $\ell_{0}\approx2\;\mbox{kpc}$, hence $\ell_{0}/L_{\mathrm{J}0}\approx1$.}

Before discussing the results of our analysis, let us make a final remark
about our parameter choice.  By fixing $Q_{0}$ while changing
$\ell_{0}/L_{\mathrm{J}0}$, we are also implicitly changing $\Sigma_{0}$ and
$\sigma_{0}$, albeit not according to a unique set of scaling relations
(otherwise $Q_{0}$ would also change).  Doing so, we are probing
\emph{different} physical states or regions of the disc.  Our parameter
choice is meant to illustrate a few interesting examples of disc stability
properties relevant to high-redshift star-forming galaxies, as we discuss
below.

\subsubsection{Effect of varying the linewidth-size scaling relation
               of the clumps}

Let us first discuss the effect of varying $b$, which is illustrated in
Fig.\ 2.  We only show the size range of unstable regions (shaded), since the
associated mass range follows trivially from Larson's third law
($M\propto\ell^{2}$; remember that here $a=0$).  For $b=0$, the classical
case of Toomre instability, the range of unstable scales can be easily
computed since $\omega^{2}(\ell)<0$ is a quadratic inequality.  The largest
and the smallest (marginally) unstable scales are then
$\ell=(8\pm4\sqrt{3}\,)L_{\mathrm{J}0}\approx15L_{\mathrm{J}0}\mbox{ and
}L_{\mathrm{J}0}$.  Within this range, there is a scale that corresponds to
the fastest growing mode, and therefore plays a primary role in the classical
instability scenario.  This is `the most unstable scale', which can be
computed by minimizing $\omega^{2}(\ell)$:
$\ell=2L_{\mathrm{J}0}\ (=2\sigma_{0}^{2}/G\Sigma_{0})$.  For all other
values of $b$, the three scales introduced above depend on the coupling
between gravitational instability and spatial resolution.

When the 2D Jeans length is not resolved, the disc is unstable over a broad
range of $\ell$ for all values of $b$ (see the left panel of Fig.\ 2).  The
largest unstable scale decreases gradually with increasing $b$, while the
smallest and the most unstable scales decrease steeply as $b$ approaches 0.5
(Regime A) and vanish thereafter (Regimes B and C).  Note, however, that
scales $\ell\la2L_{\mathrm{J}0}$ cannot be resolved, and scales
$\ell\ga8L_{\mathrm{J}0}$ are unphysical because they exceed the typical size
of galaxies at $z\sim2$.%
\footnote{SINS/zC-SINF galaxies have a median radial extent of about 11 kpc
  (see figs 2--20 of Genzel et al.\ 2014).  This is about twice the median
  half-light radius (see now their table 1), and eight times the 2D Jeans
  length (remember that $L_{\mathrm{J}0}\approx1.4\;\mbox{kpc}$).}
This implies (i) that the observable size range of unstable regions is
constant up to $b\approx0.5$, and shrinks by a factor of 6 from $b\approx0.5$
to $b=1.1$; and (ii) that none of the `characteristic' unstable scales plays
a significant role in Toomre-like instabilities, when
$\ell_{0}=2L_{\mathrm{J}0}$ and $Q_{0}=0.5$.

When the 2D Jeans length is resolved, the disc is no longer unstable for all
values of $b$ (see now the middle and right panels of Fig.\ 2).  There are
two distinct instability domains, but only one of them is observable: the
domain of Toomre-like instabilities (Regime A).  The higher the spatial
resolution, the smaller this domain.  Note that there is a value of $b<0.5$
for which the disc is marginally unstable, like a classical disc with
$Q_{0}=1$.  In such a case, the instability range collapses into a single
characteristic scale, which is of the order of the typical size of the clumps
(e.g., Noguchi 1998, 1999; Dekel et al.\ 2009; Genzel et al.\ 2011).  When
$\ell_{0}=L_{\mathrm{J}0}$, marginal instability occurs for $b\approx0.4$ and
the characteristic instability scale is about $2.5L_{\mathrm{J}0}$,
i.e.\ 25\% larger than in the classical case ($2L_{\mathrm{J}0}$).  When
$\ell_{0}=0.5L_{\mathrm{J}0}$, the disc is marginally unstable for
$b\approx0.25$ and the characteristic scale is about $5L_{\mathrm{J}0}$,
which is comparable to the half-light radius of the galaxy.

\subsubsection{Effect of varying the mass-size scaling relation
               of the clumps}

The effect of varying $a$ is illustrated in Fig.\ 3.  A comparison between
the top panels of this figure and Fig.\ 2 shows that increasing $a$ has a
qualitatively similar effect to decreasing $b$.  This is basically because,
as $a$ varies from $-$0.8 to 0.7, the disc spans all the main instability
regimes, starting from Regime C and ending with Regime A.  Despite this
similarity, $a$ has a stronger impact on disc instability than $b$.  For
example, when $\ell_{0}=L_{\mathrm{J}0}$, the characteristic instability
scale for a marginally unstable disc ($a\approx0.15$) is about
$3L_{\mathrm{J}0}$, i.e.\ 50\% larger than in the classical case.  And, when
$\ell_{0}=0.5L_{\mathrm{J}0}$, such a scale exceeds the typical size of
galaxies at $z\sim2$.  The bottom panels of Fig.\ 3 show that $a$ has an even
stronger impact on the mass range of unstable regions.  This is because
$M=M_{\mathrm{J}0}\,(\ell/L_{\mathrm{J}0})^{2+a}$ with $2+a>1$, and because
variations in $M$ are now boosted by the $a$-dependent factor
$(\ell/L_{\mathrm{J}0})^{a}$.  Note also that the 2D Jeans mass is defined
consistent with the mass-size scaling relation,
$M_{\mathrm{J}0}\equiv\Sigma(L_{\mathrm{J}0})\,L_{\mathrm{J}0}^{2}$, and so
are all other relevant masses.  Hence the lower and upper bounds of the
observable range are themselves functions of $a$ for a given
$M_{\mathrm{J}0}$, as is shown in the bottom panels of Fig.\ 3.

\subsection{Discs close to marginal instability}

Current dynamical models of high-redshift star-forming galaxies suggest that
their discs are driven by self-regulation processes, which keep them close to
marginal instability (e.g., Noguchi 1998, 1999; Agertz et al.\ 2009a; Dekel
et al.\ 2009; Burkert et al.\ 2010; Krumholz \& Burkert 2010; Cacciato et
al.\ 2012; Forbes et al.\ 2012, 2014).  In Sect.\ 2.2, we have shown that
clumpy discs can be marginally unstable even if $Q_{0}\ll1$.  Here we analyse
the case $Q_{0}=1$, which is classically associated with marginal
instability.  Note that this value of $Q_{0}$ is close to the median value
$Q_{0}\approx0.9$ found in the inner discs of SINS/zC-SINF galaxies at
$z\sim2$, as we infer from table 1 of Genzel et al.\ (2014).  Note also that
the median value of the 2D Jeans length in such discs is
$L_{\mathrm{J}0}\approx0.4\;\mbox{kpc}$, and that the spatial resolution
scale is $\ell_{0}\approx2\;\mbox{kpc}$, hence
$\ell_{0}/L_{\mathrm{J}0}\approx5$.  This means that the 2D Jeans length is
far from being resolved, and so are the size and mass ranges of unstable
regions for all observed values of $a$ and $b$.  This is consistent with the
gravitational quenching found by Genzel et al.\ (2014), but it also means
that we need much higher resolution to probe gravitational instabilities in
such discs.

What would we observe if the 2D Jeans length were marginally resolved
($\ell_{0}=L_{\mathrm{J}0}$) and the Toomre parameter were still unity?  As
$a$ and $b$ span the ranges analysed in Sect.\ 2.2, we would observe two
instability domains: the classical domain of marginally unstable discs
($a=b=0$, $\ell=2L_{\mathrm{J}0}$, $M=4M_{\mathrm{J}0}$), and a domain of
Toomre-like instabilities ($0.5\leq a\leq0.7$, $b=\frac{1}{2}$).  In such a
case, the disc is marginally unstable for $a=b=\frac{1}{2}$, the
characteristic instability scale is $\ell\approx4L_{\mathrm{J}0}$, and the
associated mass is $M\approx30M_{\mathrm{J}0}$.  This is consistent with the
results of Romeo et al.\ (2010), who found that the stability criterion for
Regime A degenerates into Toomre's stability criterion for all $a=b$.  As
$Q_{0}=1$ is a case of special interest, let us also analyse specific values
of $(a,b)$: the cases illustrated in Fig.\ 1.
\begin{itemize}
\item \emph{Case L}, i.e.\ $(a,b)=(-0.1,0.38)$, represents the original
  scaling relations found by \underline{L}arson (1981).
\item \emph{Case S}, i.e.\ $(a,b)=(0,0.50\pm0.05)$, corresponds to the
  scaling relations found by \underline{S}olomon et al.\ (1987).  Without the
  error bars, these are the well-known Larson's scaling laws.
\item \emph{Case K}, i.e.\ $(a,b)=(\frac{1}{3},\frac{1}{2})$, is the result
  of a detailed comparative analysis between observations of molecular
  clouds, high-resolution simulations and advanced models of supersonic
  turbulence (\underline{K}ritsuk et al.\ 2013; \underline{K}ritsuk, private
  communication).%
\footnote{Such scaling relations also apply to the cold atomic gas, while the
  warm component has $0.5<a<1$ and $b=\frac{1}{3}$ (Kritsuk, private
  communication).}
\item \emph{Case Fc}, i.e.\ $(a,b)=(0.44\pm0.14,0.49\pm0.02)$, is a
  prediction based on state-of-the-art simulations of supersonic turbulence
  with \underline{c}ompressive driving (\underline{F}ederrath 2013;
  \underline{F}ederrath, private communication).
\item \emph{Case Fs}, i.e.\ $(a,b)=(0.58\pm0.03,0.48\pm0.02)$: same as Case
  Fc, but for \underline{s}olenoidal driving.
\end{itemize}
As we move along this sequence of cases, $b$ remains approximately constant
and close to 0.5, while $a$ varies from $-$0.1 to 0.6.  This gives rise to
significant differences in the stability properties of the disc, especially
in its stability threshold ($\overline{Q}_{0}$) and stability level
($Q_{0}/\,\overline{Q}_{0}$), given that most of these cases fall within
Regime A.  In fact, as we move from L to Fs while keeping $Q_{0}=1$ and
$\ell_{0}=L_{\mathrm{J}0}$, the disc changes from highly stable (L and S) to
unstable (Fs).  In this case, the instability range is
$\ell\ga2L_{\mathrm{J}0}$, the most unstable scale is
$\ell\approx4L_{\mathrm{J}0}$, and the (in)stability threshold is
$\overline{Q}_{0}\approx1.1$.  This reveals an important peculiarity of
clumpy discs: they can be unstable across a wide range of scales and, at the
same time, close to marginal instability!  This is not a paradox.  It follows
from the fact that the dispersion relation of such discs can be very flat
and/or asymmetric around its minimum.

\section{ROLE OF THE DISC SCALE HEIGHT}

The mass-size and linewidth-size scaling relations considered so far are
simple power laws, like those observed in molecular clouds but extrapolated
to scales larger than the disc scale height (see Sect.\ 2.1).  There is
indeed no direct measurement of those relations at such scales.  Most of what
we know relies on the power spectra of gas and dust intensity fluctuations
observed in nearby galaxies, or related diagnostics (see, e.g., Hennebelle \&
Falgarone 2012, and references therein).  In the best-resolved cases, the
power spectrum is a double power law, with a break at scales comparable to
the disc scale height: $\ell\approx h$ (e.g., Elmegreen et al.\ 2001; Dutta
et al.\ 2009; Block et al.\ 2010; Combes et al.\ 2012).  This break is also
observed in high-resolution simulations of gas-rich galaxies (Bournaud et
al.\ 2010; Combes et al.\ 2012), and is interpreted as a transition from 3D
($\ell\la h$) to 2D ($\ell\ga h$) turbulence.  A thorough discussion of such
regimes is given by Bournaud et al.\ (2010).

It is highly non-trivial to translate observed power spectra into mass-size
or linewidth-size scaling relations.  The reason is twofold:
\begin{enumerate}
\item both density and velocity fluctuations contribute to the intensity
  power spectrum (Lazarian \& Pogosyan 2000);
\item even when density fluctuations dominate, there are distinct methods for
  estimating the fractal dimension, $D$, which lead to significantly
  different values of $a=D-2$ (Federrath et al.\ 2009; Federrath, private
  communication).
\end{enumerate}
S\'{a}nchez et al.\ (2010) carried out a detailed fractal analysis of M\,33,
and showed that the distribution of molecular gas undergoes a transition from
fractal to homogeneous at scales roughly comparable to the disc scale height.
This suggests that $a\approx0$ for $\ell\ga h$.  Concerning the value of $b$,
Kim et al.\ (2007) analysed the physical properties of atomic gas in the
Large Magellanic Cloud, and found that the linewidth-size scaling relation of
H\,\textsc{i} clouds is a simple power law up to scales of a few kpc.
Bournaud et al.\ (2010) analysed the gas velocity fields of simulated
galaxies, and found that the power spectrum of $v_{z}$ has a break at
$\ell\approx h$, while the power spectra of $v_{R}$ and $v_{\phi}$ are simple
power laws.  Since the velocity dispersion relevant to our stability analysis
is the radial one, the results above suggest that $b\approx constant$ up to
scales $\ell\gg h$.

\begin{figure}
\includegraphics[angle=-90.,scale=.97]{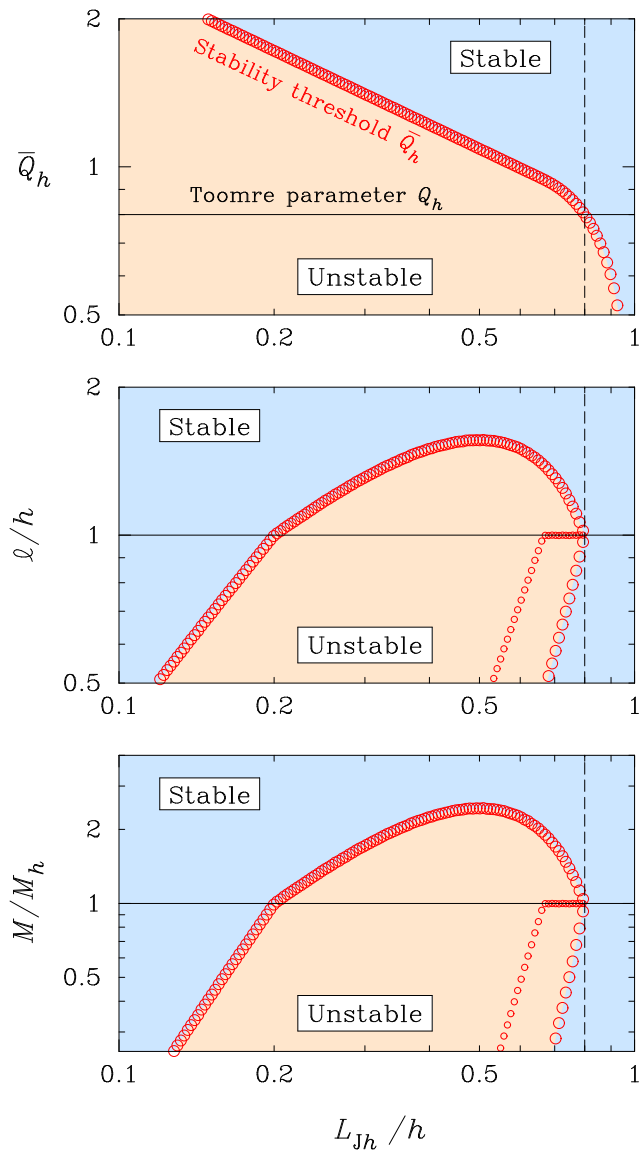}
\caption{Stability properties of clumpy discs at high redshift: effect of a
  break in the mass-size scaling relation.  The three panels show the
  stability threshold (top), the size range of unstable regions and the most
  unstable scale (middle), and the associated masses (bottom).  The clump
  scaling relations are $\sigma(\ell)=\sigma_{h}(\ell/h)^{1/2}$,
  $\Sigma(\ell)=\Sigma_{h}(\ell/h)^{1/3}$ if $\ell\leq h$ and
  $\Sigma(\ell)=\Sigma_{h}$ otherwise, where $h$ is the disc scale height.
  This is suggested by observations and simulations of galactic turbulence
  (see Sect.\ 3).  $Q_{h}\equiv\kappa\sigma_{h}/\pi G\Sigma_{h}$ and
  $L_{\mathrm{J}h}\equiv\sigma_{h}^{2}/G\Sigma_{h}$ are the Toomre parameter
  and the 2D Jeans length at scale $\ell=h$.  The case $Q_{h}=0.8$ is shown
  for illustrative purposes.  Qualitatively similar results are found for all
  values of $Q_{h}\la1$ (see again Sect.\ 3).}
\end{figure}

So what role does the disc scale height play in our stability scenario?  To
answer this question, we consider the following mass-size and linewidth-size
scaling relations:
\begin{equation}
\Sigma(\ell)=\Sigma_{h}\left(\frac{\ell}{h}\right)^{a},\;\;\;\;\;
a=\left\{\begin{array}{ll}
         1/3 & \mbox{if\ }\ell\leq h\,, \\
         0   & \mbox{else}\,;
         \end{array}
  \right.
\end{equation}
\begin{equation}
\sigma(\ell)=\sigma_{h}\left(\frac{\ell}{h}\right)^{b},\;\;\;\;\;
b=1/2\,.
\end{equation}
Such values of $a$ and $b$ are motivated by the results discussed above, and
by the detailed comparative analysis carried out by Kritsuk et al.\ (2013)
for $\ell\la h$ (see Sect.\ 2.3).  Fig.\ 4 illustrates that a break in the
mass-size scaling relation causes a transition in the stability properties of
the disc.  Look for example at the middle panel, and see how the size range
of unstable regions and the most unstable scale `break' at $\ell=h$.  Such a
transition exists for all values of $Q_{h}\la1$, and should be observable if
the disc scale height is spatially resolved.  Note also that the
characteristic instability scale for marginally unstable discs
($Q_{h}=\overline{Q}_{h}$) is
\begin{equation}
\ell_{\mathrm{c}}=h\approx L_{\mathrm{J}h}\,.
\end{equation}
This means that the disc scale height is also the natural size of unstable
clumps, and is comparable to the 2D Jeans length (for
$Q_{h}=\overline{Q}_{h}$).  Is this an obvious result?  No, it is not!  In
non-clumpy but realistically thick gas discs, the characteristic instability
scale is
\begin{equation}
\ell_{\mathrm{c}}\approx4\pi h=4L_{\mathrm{J}}
\end{equation}
(see Appendix A).  This is well beyond the ranges shown in Fig.\ 4.

The results discussed above show that the disc scale height plays an
important role in our stability scenario.  This is a promising and novel
avenue for constraining the size and mass of star-forming clumps in
high-redshift galaxies, a topic that we will address further in future work.

\section{SUMMARY AND CONCLUSIONS}

We have investigated the gravitational instability of clumpy disc
galaxies, focusing on the size and mass ranges of unstable regions.
Multi-frequency observations of both the gas and the stellar contents
(e.g., Elmegreen \& Elmegreen 2006; Shapiro et al.\ 2008; Tacconi et
al.\ 2010) have established that such galaxies are ubiquitous at high
redshift.  Furthermore, the majority of stars in the Universe are
known to form at $z>1$ (e.g., Hopkins \& Beacom 2006).  Thus it is
crucial to understand the properties of unstable star-forming gas at
this epoch of galaxy evolution.

Clumpy discs at high redshift are dynamically similar to gas discs
with scale-dependent surface density and velocity dispersion,
i.e.\ $\Sigma\propto\ell^{a}$ and $\sigma\propto\ell^{b}$, where
$\ell$ is the clump size.  Taking these `turbulent' scaling relations
into account, and extending the traditional Toomre stability analysis
as in Romeo et al.\ (2010), a wide variety of non-classical stability
properties arise.  We have illustrated this scenario for the whole
observed range spanned by the clump scaling relations, which is
centred around Larson's scaling laws $(a,b)=(0,\frac{1}{2})$, and for
a range of spatial resolution scales typical of current high-redshift
surveys.  Our key results and a few eloquent examples are summarized
below.

\begin{enumerate}
\item The scale-dependence of the surface density and velocity
  dispersion plays a \emph{crucial} role in determining the size and
  mass ranges of unstable regions.  For example, in the rings and
  outer discs of SINS/zC-SINF galaxies at $z\sim2$ (Genzel et
  al.\ 2014), where the spatial resolution scale is close to the
  inferred 2D Jeans length, small variations in the logarithmic slope
  of $\Sigma(\ell)$ can lead to dramatic, order-of-magnitude, changes
  in the mass of the most unstable clumps.  For the same observed
  surface density and velocity dispersion, logarithmic slopes of
  $\sigma(\ell)$ steeper than $b\approx0.4$ and flatter than
  $b\approx0.5$ ($a=0$) lead to complete disc stability.  This
  illustrates the dynamical complexity introduced by the clump scaling
  relations.
\item Variations in the logarithmic slopes of $\Sigma(\ell)$ and
  $\sigma(\ell)$ can drive significant changes in the stability
  properties of the disc at \emph{all} scales.  For example, a clumpy
  disc can be marginally stable even if the classical Toomre parameter
  $Q_{0}\ll1$.  In the case of Larson's scaling laws, the disc is
  always stable, however small $Q_{0}$ is, if the inferred 2D Jeans
  length $L_{\mathrm{J}0}$ is larger than the spatial resolution scale
  $\ell_{0}$.
\item For discs with $Q_{0}=1$, we have payed special attention to
  $b\approx0.5$ and $-0.1\la a\la0.6$, since this range encompasses
  the most representative values of $a$ and $b$ found in observational
  (e.g., Larson 1981; Solomon et al.\ 1987) and theoretical (e.g.,
  Federrath 2013; Kritsuk et al.\ 2013) works on supersonic
  turbulence.  In spite of being marginally stable in the classical
  sense, such discs can be anywhere from highly stable to unstable,
  depending on the value of $a$.  In fact, as $a$ approaches 0.6 while
  $L_{\mathrm{J}0}=\ell_{0}$, all observable scales
  $\ell\ga2L_{\mathrm{J}0}$ become unstable, \emph{even though} the
  disc is close to the stability threshold
  ($\overline{Q}_{0}\approx1.1$).
\end{enumerate}

Points (i)--(iii) illustrate the peculiar stability regimes possessed
by discs with scale-dependent surface densities and velocity
dispersions, and why it is important to take such regimes into account
when predicting the size and mass of star-forming clumps in
high-redshift galaxies.  Note also that our work raises an important
caveat: as the interstellar medium (ISM) is characterized by
scale-dependent surface densities and velocity dispersions, we cannot
thoroughly understand its global stability properties unless we carry
out \emph{multi-scale} observations.  This will soon be possible
thanks to dedicated ALMA surveys, which will explore the physical
properties of super-giant molecular clouds at the peak of cosmic star
formation and beyond.

Our work provides a new set of tools for exploring galactic star
formation.  In the ISM, there exist different sources of turbulence
driving, such as large-scale gravitational stirring and stellar
feedback (e.g., Mac Low \& Klessen 2004; Agertz et al.\ 2009a), and it
is still unclear how they affect the ISM at various scales.
Understanding the origin and evolution of $a$ and $b$, and how they
vary with galactic environment, is a daunting task for numerical
simulations, given the vast dynamical range involved in the
star-forming ISM: from scales $\ell\la0.1\,\mbox{pc}$ to scales
$\ell\sim10\,\mbox{kpc}$.  Preliminary results from numerical
simulations of entire galactic discs (Agertz, Romeo \& Grisdale, in
preparation) show that large-scale gravitational stirring and stellar
feedback can generate markedly different scaling properties in both
$\Sigma(\ell)$ and $\sigma(\ell)$.  This is a promising and novel
avenue for constraining the role of stellar feedback in galaxy
evolution, a topic that we will address further in future work.

\section*{ACKNOWLEDGMENTS}

We are very grateful to Reinhard Genzel for giving us more information about
the properties of rings and outer discs in SINS/zC-SINF galaxies at $z\sim2$;
and to Andreas Burkert, Sami Dib, Guillaume Drouart, Christoph Federrath,
Kirsten Kraiberg Knudsen, Alexei Kritsuk and Mathieu Puech for useful
discussions.  We are also grateful to an anonymous referee for constructive
comments and suggestions, and for encouraging future work on the topic.  ABR
thanks the warm hospitality of the Department of Fundamental Physics at
Chalmers.

\appendix

\section{DERIVATION OF EQ.\ (13)}

Consider a gas disc of scale height $h$, and perturb it with axisymmetric
waves of frequency $\omega$ and wavenumber $k$.  The response of the disc is
described by the dispersion relation
\begin{equation}
\omega^{2}=\kappa^{2}-\frac{2\pi G\Sigma\,k}{1+kh}+\sigma^{2}\,k^{2}\,,
\end{equation}
where $\kappa$ is the epicyclic frequency, $\Sigma$ is the surface density at
equilibrium, and $\sigma$ is the 1D velocity dispersion (Vandervoort 1970;
Romeo 1992, 1994; Elmegreen 2011; Griv \& Gedalin 2012 extended this analysis
to non-axisymmetric waves).  So the three terms on the right side of
Eq.\ (A1) represent the contributions of rotation, self-gravity and pressure.
For $kh\ll1$, Eq.\ (A1) reduces to the usual dispersion relation for an
infinitesimally thin gas disc.  For $kh\gg1$, one recovers the case of Jeans
instability with rotation, since $\Sigma/h=2\rho$.  In other words, scales
comparable to $h$ mark the transition from 2D to 3D stability.

If the disc is self-gravitating and isothermal along the vertical direction,
as assumed in the analyses above, then the disc scale height is closely
related to the 2D Jeans length:
\begin{equation}
h=\frac{\sigma^{2}}{\pi G\Sigma}=\frac{L_{\mathrm{J}}}{\pi}\,.
\end{equation}
To compute the characteristic instability scale, we express the dispersion
relation in a form similar to Eq.\ (9):
\begin{equation}
\frac{\omega^{2}}{\kappa^{2}}=1-
\frac{4}{Q^{2}}\,
\frac{1}{2+(\ell/L_{\mathrm{J}})}+
\frac{4}{Q^{2}}\,
\frac{1}{(\ell/L_{\mathrm{J}})^{2}}\,.
\end{equation}
The most unstable scale is the scale that minimizes the dispersion relation:
$\ell_{\mathrm{min}}\approx4L_{\mathrm{J}}$.  In this classical case,
$\ell_{\mathrm{min}}$ does not depend on whether the disc is marginally
unstable or not, and is therefore the characteristic instability scale:
\begin{equation}
\ell_{\mathrm{c}}=\ell_{\mathrm{min}}\approx\frac{4\sigma^{2}}{G\Sigma}\,.
\end{equation}
Using Eq.\ (A2), we find that
\begin{equation}
\ell_{\mathrm{c}}\approx4\pi h=4L_{\mathrm{J}}\,,
\end{equation}
which is Eq.\ (13) of the main text.

\bsp

\label{lastpage}

\end{document}